\documentclass[%
 reprint,
 amsmath,amssymb,
 aps,
]{revtex4-2}

\usepackage{graphicx}
\usepackage{dcolumn}
\usepackage{bm}
\usepackage{amsmath}	
\usepackage{amssymb}	
\usepackage{physics}
\usepackage{tabularx}
\usepackage{makecell}
\usepackage{booktabs}
\usepackage[dvipsnames]{xcolor}

\newcommand{\Pxshot}{P_\times^{\mathrm{shot}}}
\newcommand{\Pshot}{P^{\mathrm{shot}}}
\newcommand{\fnl}{f_{\mathrm{NL}}}
\newcommand{\sigmafnl}{\sigma_{\fnl}}
\newcommand{\Msun}{\textup{M}_\odot}
\newcommand{\be}{\begin{equation}}
\newcommand{\ee}{\end{equation}}
\newcommand{\sun}{\odot}
\newcommand\ion[2]{\text{#1\,\textsc{\lowercase{#2}}}}	

\newcommand\apjl{ApJ}                
\newcommand\baas{BAAS}               
\newcommand\jcap{J.~Cosmology Astropart. Phys.} 

\newcommand\mnras{MNRAS}             

\newcommand\physrep{Phys.~Rep.}      

\newcommand{\pcb}[1]{{\color{black} #1}}

\begin{document}

\preprint{APS/123-QED}

\title{Coupling parsec and gigaparsec scales: primordial non-Gaussianity with multi-tracer intensity mapping}

\author{R. Henry Liu$^{1,\,2}$, }
 \email{\href{mailto:rhliu@phas.ubc.ca}{rhliu@phas.ubc.ca}}
\author{Patrick C. Breysse$^{2,\,3}$}%
\affiliation{%
 $^{1}$Department of Physics and Astronomy, University of British Columbia, 6224 Agricultural Road, Vancouver, V6T 1Z1, Canada \\
 $^{2}$Canadian Institute for Theoretical Astrophysics, University of Toronto, 60 St. George Street, Toronto, ON, M5S 3H8, Canada \\
 $^{3}$Center for Cosmology and Particle Physics, Department of physics, New York University, 726 Broadway, New York, NY, 10003, U.S.A.
}%


\date{\today}

\begin{abstract}
Primordial non-Gaussianity (PNG) is a key probe of the origins of primordial fluctuations in the early universe. It has been shown that multi-tracer measurements of large-scale structure can produce high-precision measurements of PNG.  Future line intensity mapping surveys are well-suited to these measurements owing to their ability to rapidly survey large volumes and access the large scales at which PNG becomes important.  In this paper, we explore for the first time how multi-tracer PNG measurements with intensity mapping surveys depend on the sub-galactic scale physics which drives line emission.  We consider an example cross-correlation between CO maps, and quantify the impact varying the astrophysical model has on $\fnl$ measurements.  We find a non-trivial coupling between horizon-scale PNG measurements and the molecular cloud-scale interstellar medium that can have order unity effects on $\fnl$ constraints with near-future experimental sensitivities. We discuss how these effects depend on noise level and survey design.  We further find that the cross-correlation shot noise, an effect nearly unique to intensity mapping measurements, can play an important role in multi-tracer analyses and should not be neglected.
\end{abstract}

\maketitle


\section{Introduction}
    
In the current concordance model of cosmology, the earliest phase of cosmic history is an epoch of rapid inflation, when the Universe expanded much more quickly than any time after.  In addition to explaining the observed flatness and homogeneity of the Universe, the inflationary paradigm also provides a source for the primordial density perturbations which sourced its large-scale structure.  However, despite its utility as a model, the actual mechanism behind inflation remains unknown.  

One key observable which can be used to distinguish between different inflation models is primordial non-Gaussianity (PNG).  The simplest models of cosmic inflation predict a Gaussian primordial density field, so any measurement of non-Gaussian initial conditions provides a powerful discriminant between early universe models (see, e.g., \citealt{Bartolo2004,Chen2010}).  We can quantify deviations from Gaussianity using a shape function and an amplitude $\fnl$.  In this work we will examine local-type PNG \citep{Gangui1994,Verde2000,Wang2000,Komatsu2001}, which measures correlations between very long- and short-wavelength density modes.  This specific shape is a powerful probe of multi-field inflation models, as single-field models predict very little PNG of this type \citep{Maldacena2003,Creminelli2004}.  Currently, the best constraints on local PNG come from via the \textit{Planck} satellite \citep{Planck2018PNG}. \textit{Planck} was able to produce a constraint of $f_{\mathrm{NL}}^{\mathrm{local}}=-0.9\pm 5.1$ from the latest CMB temperature and polarization measurements. This result is consistent with mostly Gaussian primordial fluctuations, ($\fnl = 0$). However, greater accuracy even than this is desired in order to better constrain the behavior of PNG, with $\sigma_{\fnl}\sim1$ often cited as a useful target (see, e.g. Ref~\cite{Meerburg2019} and references therein).


To go beyond the \textit{Planck} constraints, we can use a specific behavior  that PNG is known to produce in the two-point statistics of biased large-scale structure tracers.  Local-type PNG produces a strongly scale dependent bias on very large scales \citep{Dalal2008, Matarrese2008, Slosar2008}. Typically, this feature is sought after in large galaxy surveys \citep[see, e.g.][]{Slosar2008,Camera2013, Camera2015, Raccanelli2015, Alonso2015,Amendola2018}.  Though this signature provides a distinctive probe of primordial physics, the fact that it only appears on very large scales makes it challenging to measure in practice.  The finite volume of the observable universe limits the number of density modes available on very large scales.  

In this paper we combine two methods which have been individually proposed for accessing this large-scale information.  The first is multi-tracer analysis \citep{Seljak2009,Mcdonald2009,Hamaus2011,Abramo2013}. It has been shown that, though measurements of a single tracer are dominated by cosmic variance on large scales, cross-correlations between tracers with different biases face no such limitation.  In a single-tracer analysis, the error budget is limited by the number of modes available, which becomes relatively small at large scales.  With two tracers, however, there is additional information in each individual mode which can be arbitrarily increased by improving signal to noise.

The second method we will examine is line intensity mapping (LIM) \citep{KovetzLIM2017}. Noise in a galaxy survey is set by how many galaxies can be detected, so a high-quality $\fnl$ measurement requires directly imaging large numbers of sources over huge volumes.  LIM provides an alternative method which can map large areas with less onerous sensitivity requirements.  Rather than observing individual sources, LIM surveys map the large-scale fluctuations in the intensity of a chosen emission line, obtaining three-dimensional information by observing at may closely-spaced frequencies.  In this way, LIM is sensitive to the aggregate emission from all sources within a given volume.  LIM surveys, with their ability to access large scales, have been shown to provide significant constraining power on $\fnl$ \citep{Camera2013,Camera2019,Dizgah2019a,Dizgah2019b}.

Recent years have seen the development of a large number of intensity mapping experiments targeting several different emission lines.  The most commonly targeted line is the 21 cm hyperfine line emitted by neutral hydrogen \citep{Pritchard2012}, which has been successfully mapped in cross-correlation with galaxy surveys by the Green Bank \citep{Chang2010,Masui2013,Switzer2013} and Parkes \citep{Anderson2018} telescopes.  Recently, a number of experiments have been developed targeting other lines, including rotational transitions of carbon monoxide molecules \cite{Pullen2013,Breysse2014,Li2016,Keating2016}, the bright 158 $\mu$m \ion{C}{ii} fine-structure line \citep{Gong2012,Uzgil2014,Silva2015,Chung2018,Dumitru2019,Padmanabhan2019,Yue2019}, and several others \citep{Pullen2014,Visbal2015,Comaschi2016,Fonseca2017,Gong2017,Silva2018,Sun2019}.  For an overview of the current experimental landscape, see \citet{Kovetz2019} and references therein.

It has been widely demonstrated in the literature that many of the science goals of LIM surveys are substantially enhanced through the use of cross-correlations, whether between intensity maps of different lines or between intensity maps and other observables such as galaxy surveys.  In the near term, cross-correlations can increase detection significance and help with signal validation and foreground removal \citep{Masui2013,Croft2016,Chung2019a,Yang2019,Ade2020}.  On top of this, it has been shown that cross-correlations can add information inaccessible to either individual tracer, most notably about the interstellar medium (ISM) of emitting galaxies \citep{Serra2016,Switzer2017,Wolz2017,Breysse2017,Beane2019,Breysse2019,Wolz2019} and their surrounding intergalactic medium \citep{Lidz2011,Feng2017,Heneka2017,Fialkov2020}.  The specific case we focus on here, that of multi-tracer PNG measurements, was studied by \citet{Fonseca2018} for a hypothetical correlation between intensity maps of the Ly$\alpha$ and H$\alpha$ lines.  This calculation, however, relied only on a single model for each of the target lines, effectively fixing the ISM conditions in their model.  Here we will more broadly study the significant impact that sub-galactic line emission physics can have on PNG measurements with LIM data.

One can immediately see a potential problem in seeking to combine LIM measurements with multi-tracer methods:  Since LIM surveys by definition trace every source in a target population, how is it possible to obtain two populations with different biases?  The answer comes from the fact that, by mapping line intensities, LIM surveys weight each galaxy by its line luminosity.  This means that, if we have intensity maps of two lines where different populations are bright in each line, we can benefit from the multi-tracer cosmic variance cancellation.  There is therefore a coupling between the \pcb{poorly-understood} sub-galactic-scale physics which drives line luminosity and our ability to measure the horizon-scale imprint of PNG.

\pcb{Most obviously, there is a risk that signatures of PNG could be degenerate with ISM physics.  Fortunately, this degeneracy has been found to be small for LIM observations \citep{Camera2019,Bernal2019}, as the two effects enter the observable in substantially different ways.  Here we explore a more subtle impact which has not yet been studied quantitatively.  It is well-known that multi-tracer PNG measurements are most effective when the two tracers have very different biases, but for LIM measurements the bias difference depends on the underlying astrophysical model.  That same model is also intimately coupled to other aspects of the LIM measurement, most notably the amount of shot noise.  It is therefore nontrivial to predict the amount of extra information which can be obtained from a multitracer LIM measurement.}

One should note that when we speak here of ``gaining information" by exploiting different biases that, unlike galaxy surveys, we do not have the ability to tune the bias of our sample by selecting different subpopulations.  The same wholistic nature of LIM measurements means that each line at each redshift has a fixed, unchangeable bias.  However, this bias is massively uncertain at our current level of understanding \citep[see, e.g.][]{Breysse2014,Chung2018}.  LIM observations are currently at a relatively young stage, with a wide variety of experiments targeting many different lines over disparate parts of the sky.  There is therefore a need to identify which lines and which cross-correlations offer the greatest opportunities for new science.  In this paper, we will explore the interplay between ISM physics, bias, and PNG for a simple model of a correlation between a pair of CO rotational transitions to see how PNG constraints depend on sub-galactic physics.  As we obtain more LIM detections of various lines, we will be able to use these insights to direct observations to new cross-correlation possibilities for the next generation of experiments.

We find a number of ways in which the interaction between galactic- and cosmological-scale physics in LIM surveys differs from that in conventional surveys.  \pcb{As the primary benefit of multitracer PNG measurements comes from sample variance cancellation, one does need reasonably high signal-to-noise (SNR) to see an impact.  For reasonable near-future measurements, we find that varying the luminosity-weighted bias for one line in a multi-tracer measurement can improve measurements of $\fnl$ by $\sim50\%$ or more.}  Another key effect comes from the unique way shot noise, the error due to the finite number of line emitters, appears in intensity maps.  In galaxy surveys, shot noise simply scales with the number of galaxies surveyed.  In LIM, however, we always are sensitive to every galaxy, so shot noise is fixed and determined by the same astrophysics which sets the bias.  We find that this changes the error obtained on $\fnl$ in nontrivial ways that depend sensitively on the underlying astrophysical model.  Finally, in multi-tracer LIM, every galaxy by definition appears in both surveys, leading to a non-negligible shot noise term in the cross-spectrum not present in galaxy surveys.  This effect turns out to be subdominant in our example model, but it is potentially very important for measurements with more futuristic noise levels.

The structure of this paper is as follows: In Section \ref{sec:Formalism}, we discuss the formal mathematical approach to the multi-tracer LIM problem and the resulting key effects. We then further our analysis with an example carbon monoxide experiment, which we describe in Section \ref{sec:Example_Model}. We present the results of our test model in Section \ref{sec:Results}, and discuss the potential scientific implications in Section \ref{sec:Discussion}. We then conclude this work in Section \ref{sec:Conclusions}.

\section{Formalism}
\label{sec:Formalism}

We begin by examining formally how the two-point statistics of intensity maps differ qualitatively from the more familiar galaxy survey case.

\subsection{Galaxy Surveys}
As mentioned above, the signature we seek to measure is an excess in the power spectrum of some large-scale structure tracer on very large scales.  The typical tracers used for this measurement are samples of directly detected galaxies.  Maps are made of the density contrast $\delta_g(\mathbf{x})\equiv\left(n(\mathbf{x})-\overline{n}\right)/\overline{n}$, where $n(\mathbf{x})$ is the comoving number density of the galaxy sample as a function of position.  Galaxies represent a biased tracer of large scale structure, in that
\be
\delta_g=b\delta_m,
\ee
where $\delta_m$ is the density contrast of dark matter \citep[see][and references therein for a review]{Desjacques2018}.

The power spectrum of such a biased tracer can be written as
\be
P_g(k) \equiv V\left<\left|\tilde{\delta}\right|^2(\mathbf{k})\right>= b^2(k)P_m(k)+\frac{1}{\overline{n}},
\ee
where the tilde denotes a Fourier transform, $V$ is the survey volume, $\overline{n}$ is the mean number density of galaxies in a given survey area, and $\frac{1}{\overline{n}}$ is the Poisson shot noise caused by the randomness in the galaxy positions.  The underlying dark matter density field $\delta_m$ is assumed to have power spectrum $P_m$.

For linear scales and Gaussian initial conditions, the bias $b(k)$ is scale independent.  However, as stated above, non-zero PNG will leave a correction to the halo bias, introducing a scale-dependent term on large scales.  We can write the galaxy bias as
\be
b(k)=b^0+\fnl\Delta b(k),
\ee
where $b^0$ is the scale-independent linear bias and $\fnl$ is the parameter which sets the strength of the non-Gaussianity.  For local PNG, we have

\begin{equation}
\label{eq:deltabias}
    \Delta b(k) = \frac{3(b-1)\delta_c\Omega_m H_0^2}{c^2k^2T(k)D(z)}
\end{equation}
with $\delta_c=1.686$ as the spherical collapse threshold at $z=0$, while $H_0$ is the Hubble parameter, $c$ is the speed of light, $T(k)$ is the normalized matter transfer function and $D(z)$ is the linear growth factor normalized to be $D(0)=1$ \citep{Dalal2008, Matarrese2008, Slosar2008,Afshordi2008}.

The $k^{-2}$ dependence in Eq. (\ref{eq:deltabias}) means that the dominant impact of local PNG in the power spectrum comes at the largest scales.  However, these scales can be difficult to access as these large-scale modes will typically be dominated by cosmic variance.  It has been shown in the literature, however, that the cross-correlation of two tracer populations with different biases can suppress this cosmic variance effect and improve $\fnl$ measurements.  If we have two independent galaxy samples with biases $b_1$ and $b_2$, the cross-spectrum is
\begin{equation}
    P_\times(k) = V \langle \tilde{\delta}_1(\mathbf{k}) \tilde{\delta}_2^*(\mathbf{k}) \rangle = b_1(k)b_2(k)P_m(k). 
\end{equation}
Shot noise can be thought of the correlation of each galaxy with itself, so as long as the two samples are entirely disjoint there is no shot noise in the cross-spectrum.

Combining the auto- and cross-spectra, we can compute the covariance matrix for the two correlated density fields $\delta_1$ and $\delta_2$:
\begin{align}
\label{eq:Galaxy_covariance_matrix}
\begin{split}
    C_g(k) &= 
    \begin{pmatrix} 
    P_1(k) & P_\times(k) \\
    P_\times(k) & P_2(k) 
    \end{pmatrix} \\
    &= \begin{pmatrix} 
        b_1^2 P_m + \dfrac{1}{\overline{n}_1} & 
        b_1b_2 P_m  \\
        b_1b_2 P_m  & 
        b_2^2 P_m + \dfrac{1}{\overline{n}_2} 
        \end{pmatrix},
\end{split}
\end{align}
where we have suppressed the $k$ dependence in the second equality for readability.

\subsection{Intensity Mapping}
\label{sec:Intensity_Mapping}
Because intensity maps weight each galaxy by its line luminosity, the picture becomes slightly more complicated.  We will now discuss the multi-tracer method for intensity maps. 

Consider a map of line intensity $T_1(\mathbf{x})$ in brightness Temperature units  The auto power spectrum of such a map is given by
\begin{equation}
\label{eq:LIMpowerspectrum}
    P(k) = \overline{T}_{1}^2\overline{b}_1^2(k) P_{m}(k) + P^{\mathrm{shot}}.
\end{equation}
The entire spectrum is weighted by the sky-averaged intensity of each line $\overline{T}$, which we can express as
\begin{equation}
    \overline{T} = C_{LT} \int L(M) \dv{n}{M} \dd M
\end{equation}
where $dn/dM$ is the halo mass function \citep{Tinker2008} we have assumed that a halo of mass $M$ has a line luminosity $L(M)$. $C_{LT}$ is the conversion factor between luminosity density and observed intensity for sources at redshift $z$, given by
\begin{equation}
\label{eq:CLT}
    C_{LT} = \dfrac{c^3(1+z)^2}{8\pi k_B \nu H(z)},
\end{equation}
for surveys which use brightness temperature units and
\be
C_{LT}=\frac{c}{4\pi\nu H(z)}
\ee
for those which use flux units, where $c$ is the speed of light, $k_B$ is Boltzmann's constant, $\nu$ is the rest frequency of the target line, and $H(z)$ is the Hubble parameter at redshift $z$.

We are still mapping biased tracers, but now the contribution of each source to the average bias is weighted by $L(M)$, which gives
\begin{equation}
\label{eq:LIMbias}
    \overline{b}^0 = \dfrac{\int L(M)\,b(M) \dv{n}{M} \dd M}{\int L(M) \dv{n}{M} \dd M}
\end{equation}
where $b(M)$ is the halo bias as a function of mass \citep{Tinker2010}. The $b(M)$ factor is modified in PNG by the same scale dependent modification term as the galaxy survey case, defined in Equation \ref{eq:deltabias}. 

Lastly, $P^{\mathrm{shot}}$, the Poisson noise contribution in the power spectrum is written as:
\begin{equation}
\label{eq:Pshot}
    P^{\mathrm{shot}} = C_{LT}^2 \int L(M)^2 \dv{n}{M} dM
\end{equation}
We also note that although the Poisson noise component in a galaxy survey can be reduced by increasing the number density of surveyed galaxies, the Poisson shot noise component is irreducible in the case of intensity mapping as all line emission sources are already included in an intensity map.  For a detailed derivation of the intensity mapping auto-spectrum, see the appendix of \citet{Breysse2019}.

Just as in the galaxy survey case, we can perform a multi-tracer analysis by cross-correlating two intensity maps. For the convenience of the reader, a detailed derivation of the cross spectrum between two intensity maps is provided in Appendix \ref{app:Px}, and can be expressed as:
\begin{equation}
\label{eq:Px}
    P_{\times}(k)=\overline{T}_1\overline{T}_2\overline{b}_1(k)\overline{b}_2(k)P_m(k) + P^{\mathrm{shot}}_{\times}
\end{equation}
Note the additional cross-shot term $\Pxshot$. This term originates due to the self-correlation of sources in both intensity maps. Unlike the galaxy survey case, intensity maps take contribution from all sources in the observed field, and both tracers will contain emission from the same sources, yielding a term given by,
\begin{equation}
\label{eq:Pxshot}
    P^{\mathrm{shot}}_\times = s_\times C_{LT,1}C_{LT,2}\int L_1(M)\,L_2(M)\dv{n}{M} \dd M
\end{equation}

\begin{figure}
    \centering
    \includegraphics[width=\columnwidth]{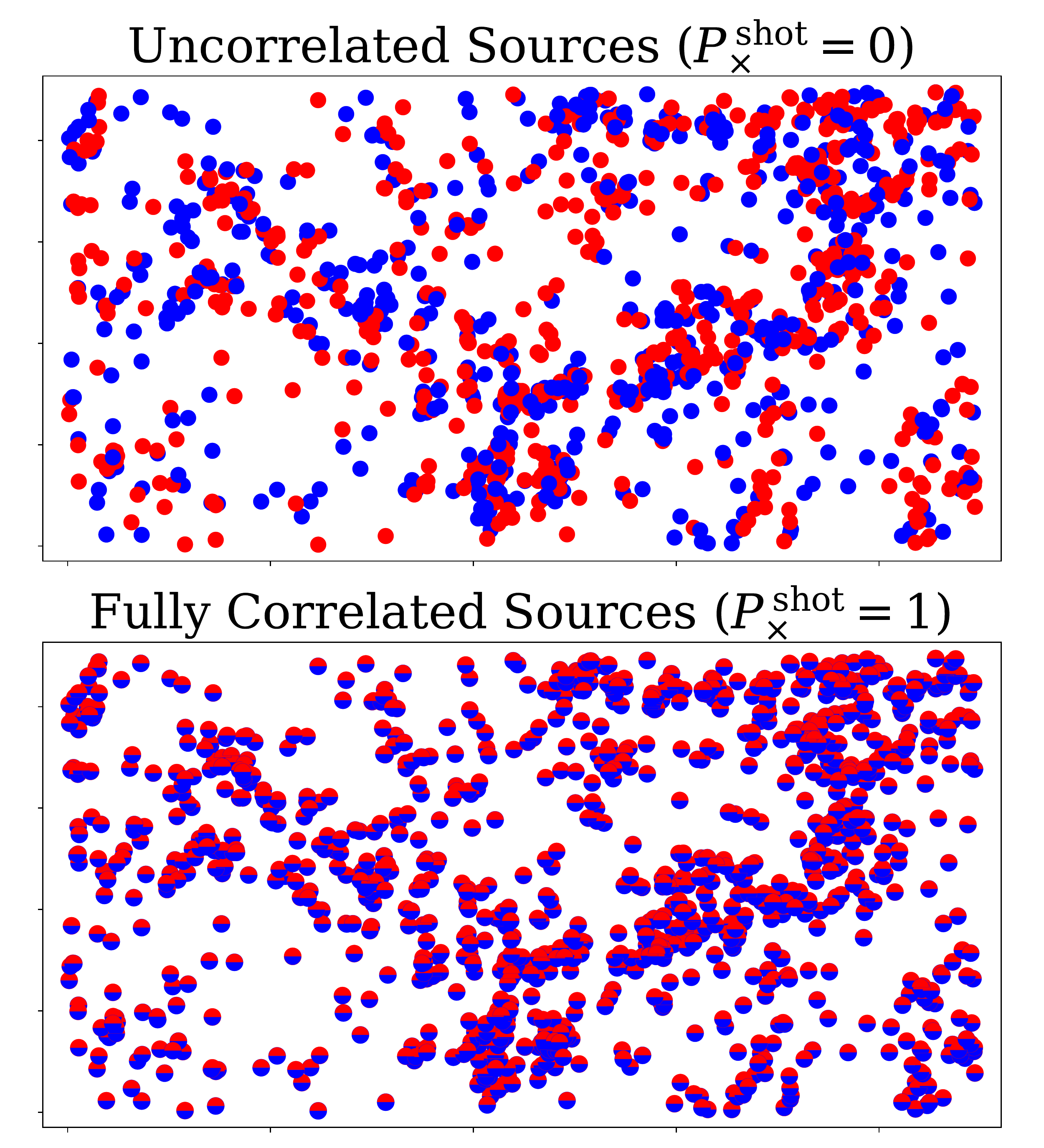}
    \caption{Simulated galaxy populations illustrating the two extreme cases of $s_\times=0$ (top) and $s_\times=1$ (bottom). In the top panel, galaxies which emit in one line do not emit in the other, and there is no correlated shot power.  In the bottom panel, every galaxy emits in both lines, and there is significant correlated shot noise.}
    \label{fig:correlated_sources}
\end{figure}

Up to this point, we have assumed that the line luminosity of a halo is entirely determined by its halo mass.  In reality, we expect there to be some scatter around an average $L(M)$ relation \citep{Li2016}.  For simplicity, we will neglect this effect with one exception.  If we allow line luminosities to have some stochasticity, then it is possible for the luminosities of two different tracer lines to scatter in different directions, which will have the effect of decreasing the correlated shot noise.  In the most extreme limit where sources bright in one line are always faint in the other, we approach the $\Pxshot=0$ limit for completely independent populations (see Figure \ref{fig:correlated_sources} for an illustration).  To account for this effect, we include a constant $s_\times$ which ranges from unity in the case of deterministic $L(M)$ relations to zero in the highly-scattered disjoint case.

With the auto and cross spectra defined, we can explicitly write the covariance matrix of two intensity fields,
\begin{align}
\label{eq:covariance_matrix}
\begin{split}
    &C(k) = 
    \begin{pmatrix} 
    P_1(k) & P_\times(k) \\
    P_\times(k) & P_2(k) 
    \end{pmatrix} \\
    &= \begin{pmatrix} 
        \overline{T}_{1}^2b_1^2 P_{m} + P_{1}^{\mathrm{shot}} & 
        \overline{T}_1\overline{T}_2b_1b_2 P_m + P^{\mathrm{shot}}_{\times} \\
        \overline{T}_1\overline{T}_2b_1b_2 P_m + P^{\mathrm{shot}}_{\times} & 
        \overline{T}_{2}^2b_2^2 P_{m} + P_{2}^{\mathrm{shot}}
        \end{pmatrix} 
\end{split}
\end{align}
With the auto and cross shot power spectra defined in Equations \ref{eq:Pshot} and \ref{eq:Pxshot}. 

We need then to add the final key ingredient which distinguishes intensity maps from typical galaxy surveys, which is noise.  We will assume here that both surveys have thermal white noise throughout, and that the noise is uncorrelated between the two surveys.  This gives a noise covariance matrix
\begin{equation}
\label{eq:noise_matrix}
    N = 
    \begin{pmatrix} 
    P_{N1} & 0 \\
    0 & P_{N2} 
    \end{pmatrix},
\end{equation}
where $P_N$ is the noise power spectrum of an intensity mapping survey.  We leave for future work any discussion of non-Gaussian noise contributions, including any effect of foreground contamination.

\subsection{Fisher Forecasts}

We utilise the Fisher formalism to estimate how well the multi-tracer approach can measure $\sigma_{\fnl}$. 
For a set of parameters $\theta_i$, the Fisher matrix at wavenumber $k$ is:
\begin{equation}
    F_{\theta_i, \theta_j}(k) = \dfrac{1}{2} \mathrm{Tr}\bigg[C_{,\theta_i}(k)(C(k)+N)^{-1}C_{,\theta_j}(k)(C(k)+N)^{-1} \bigg]
    \label{eq:fnl_k}
\end{equation}
Where $C$ and $N$ are the signal and noise covariance matrices, defined in Equations \ref{eq:covariance_matrix} and \ref{eq:noise_matrix}.  The notation $C_{,\theta}$ denotes the derivative of $C$ with respect to parameter $\theta$.  The total Fisher information is then given by summing over $k$ bins:
\be
F_{\theta_i,\theta_j}=\sum_k N_m(k)F_{\theta_i,\theta_j}(k),
\ee
where
\be
N_m(k) = \frac{k^2\Delta k V_{\rm{surv}}}{2\pi^2},
\ee
is the number of independent Fourier modes in a bin centered at $k$ with width $\Delta k$ for a survey spanning comoving volume $V_{\rm{surv}}$.

We are concerned here with exploring the effects of our multi-tracer intensity mapping analysis specifically on $\fnl$.  \pcb{It has been shown in \citet{Camera2019} and \citet{Bernal2019} that PNG measurements are not particularly degenerate with the uncertain underlying astrophysics}.  We will therefore only forecast errors on $\fnl$, assuming all other model parameters are known.  The error we obtain is then simply
\be
\sigma_{\fnl} = \frac{1}{F_{\fnl\fnl}^{1/2}}
\label{eq:sigma_fnl}
\ee
We can see from Eq. (\ref{eq:fnl_k}-\ref{eq:sigma_fnl}) that the differences between intensity mapping and galaxy survey analysis for our purposes are encoded entirely in the differences between the covariance matrices in Eqs. (\ref{eq:Galaxy_covariance_matrix}) and (\ref{eq:covariance_matrix},\ref{eq:noise_matrix}).  Specifically, we identify three key new effects in the intensity mapping case which have not been considered before:

\begin{itemize}
\item \textbf{Dependence on $\mathbf{L(M)}$-} In the conventional multi-tracer case, one typically selects a population of galaxies with relatively comparable biases.  In the LIM case, we select all galaxies by definition and any difference in bias comes from differences in $L(M)$ for the two different lines.  If the two lines have the same $L(M)$ distribution, then they will have the same biases, and we lose the cosmic variance benefits of the multi-tracer analysis (see Appendix \ref{app:same_bias} for more information).  The sub-galactic physics which sets $L(M)$ also determines the level of shot noise in a survey, complicating the effects on $\fnl$.

\item \textbf{Presence of Instrumental Noise-} The noise level in a galaxy survey sets the total number of sources which can be detected, but doesn't make any additional contribution to the power spectrum beyond that.  This extra term will be sensitive to hardware and survey design factors.

\item \textbf{Cross-shot noise-} As noted above, the cross-spectra between intensity maps will generically have a nonzero shot noise term which is not present in the galaxy survey case.  This changes the degeneracy structure of $C$ significantly, and provides an additional path by which sub-galactic astrophysics can affect $\fnl$ measurements.  \pcb{This term is subdominant in our example model at near-future instrumental sensitivities, but it has large, somewhat counterintuitive effects in the sample variance limit.  This combined with its relative lack of previous study suggests that care should be taken to include this cross-shot term in future calculations.}
\end{itemize}

\section{Example Model}
\label{sec:Example_Model}

It can clearly be seen from Section \ref{sec:Formalism} that the $\sigma_{\fnl}$ obtainable by a multi-tracer LIM analysis strongly depends both on the emission models assumed for the two lines and on the exact characteristics of the two surveys.  Even the overall amplitude of high-redshift line emission is highly uncertain (see, e.g. \citealt{Breysse2014,Chung2018}), let alone how it is distributed between haloes of different masses.  Given this lack of knowledge, it is currently difficult to predict which lines will have substantially different average biases.  On top of that, the unique impacts of multi-tracer sample-variance cancellation only come into play at high signal-to-noise, meaning that any quantitative discussion of the effects described above will likely only be relevant for future LIM experiments.  We will therefore not attempt here to fully explore the vast parameter space covered by these two uncertainties, but will focus on a single representative example of a multi-tracer LIM forecast.  The qualitative results we obtain here can be used to gain intuition about future analyses as models become more refined and instruments grow more sensitive.

Our goal is to find a pair of lines for which $L(M)$ differs in shape, not just in amplitude.  For our example, we will focus on a pair of rotational transitions of the CO molecule.  There is evidence that the slope of the CO/FIR luminosity ratio varies substantially for different CO transitions \citep{Greve2014}, which would lead to different $L(M)$ shapes.  Specifically, we choose to study maps of the 115 GHz CO(1-0) line and the 920 GHz CO(8-7) line.  Currently, the COMAP experiment is observing at $\nu_{\rm{obs}}\sim30$ GHz, and several surveys including CCAT-prime, TIME, and CONCERTO target $\nu_{\rm{obs}}\sim250$ GHZ.  With these observing frequencies, CO(1-0) and (8-7) will overlap at $z\sim3$.  It is reasonable to assume that these frequency bands will continue to be targeted by future experiments.  We will therefore examine hypothetical future versions of COMAP and CCAT-prime and study how $\sigma_{\fnl}$ depends on our choice of instrument and model parameters.

\subsection{CO Models}
\label{sec:LIM_Emission_Models}

\begin{figure}
    \centering
    \includegraphics[width=\columnwidth]{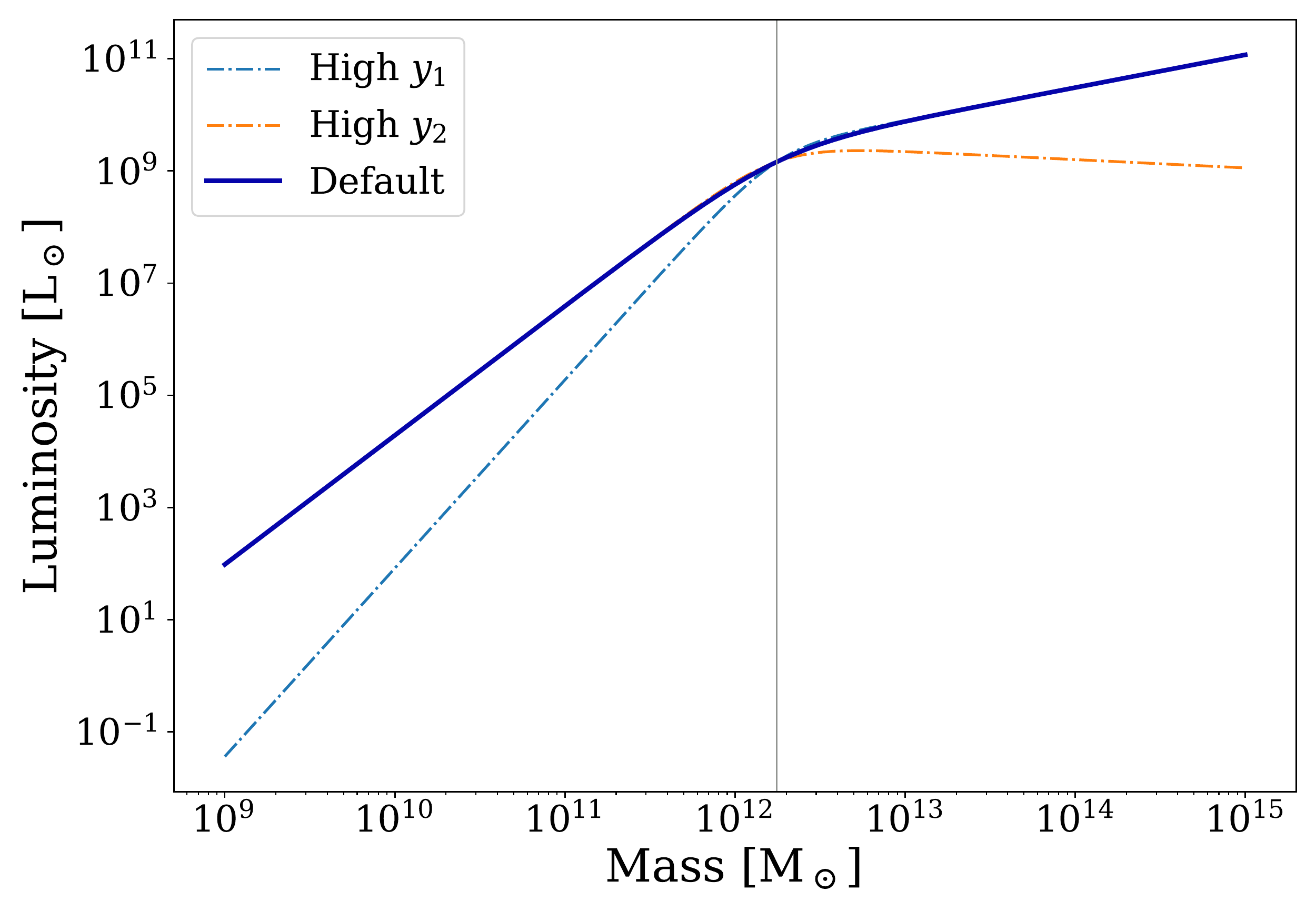}
    \caption{Double power-law mass-luminosity model given in Equation \ref{eq:doublepowerlaw} with default parameters and high $y_1$/$y_2$ parameters. The vertical grey line represents the turnover mass $M_1$.}
    \label{fig:luminositymodel}
\end{figure}

In order to forecast $\sigma_{\fnl}$ for our CO surveys, we need to assume a form of L(M) for the two lines.  For our demonstration,  we adopt the double power-law mass-luminosity model from \citet{Padmanabhan2018}:
\begin{equation}
\label{eq:doublepowerlaw}
    L(M, z) = 2A(z) M \bigg[\Big(\frac{M}{M_1(z)}\Big)^{-y_1(z)} + \Big(\frac{M}{M_1(z)}\Big)^{y_2(z)} \bigg]^{-1}
\end{equation}
We note the slight change in notation compared to \citet{Padmanabhan2018}, to avoid confusion with the bias parameter $b$ and noise covariance matrix $N$. The free parameters amplitude $A(z)$, turnover mass $M_1(z)$, low mass slope $y_1(z)$ and high mass slope $y_2(z)$ each carry redshift dependence given by
\begin{align}
    A(z) &= A_{10} + A_{11}\frac{z}{z+1} \\
    \log{M_1(z)} &= \log{M_{10}} + M_{11}\frac{z}{z+1} \\
    y_1(z) &= y_{1,\,10} + y_{1,\,11}\frac{z}{z+1} \\
    y_2(z) &= y_{2,\,10} + y_{2,\,11}\frac{z}{z+1}
    \label{eq:Luminosityparams}
\end{align}
The fiducial values for the free parameters described are given in Table \ref{tab:doublepowerlaw}, with the fitting method used to define the parameters described in \citet{Padmanabhan2018}. The resulting mass-luminosity model is plotted in Figure \ref{fig:luminositymodel}, along with the effects of varying $y_1$ and $y_2$ parameters.  We assume a minimum halo mass $M_{\mathrm{min}}=10^9\ M_{\sun}$, which is consistent with other literature models.

\begin{figure}
    \centering
    \includegraphics[width=\columnwidth]{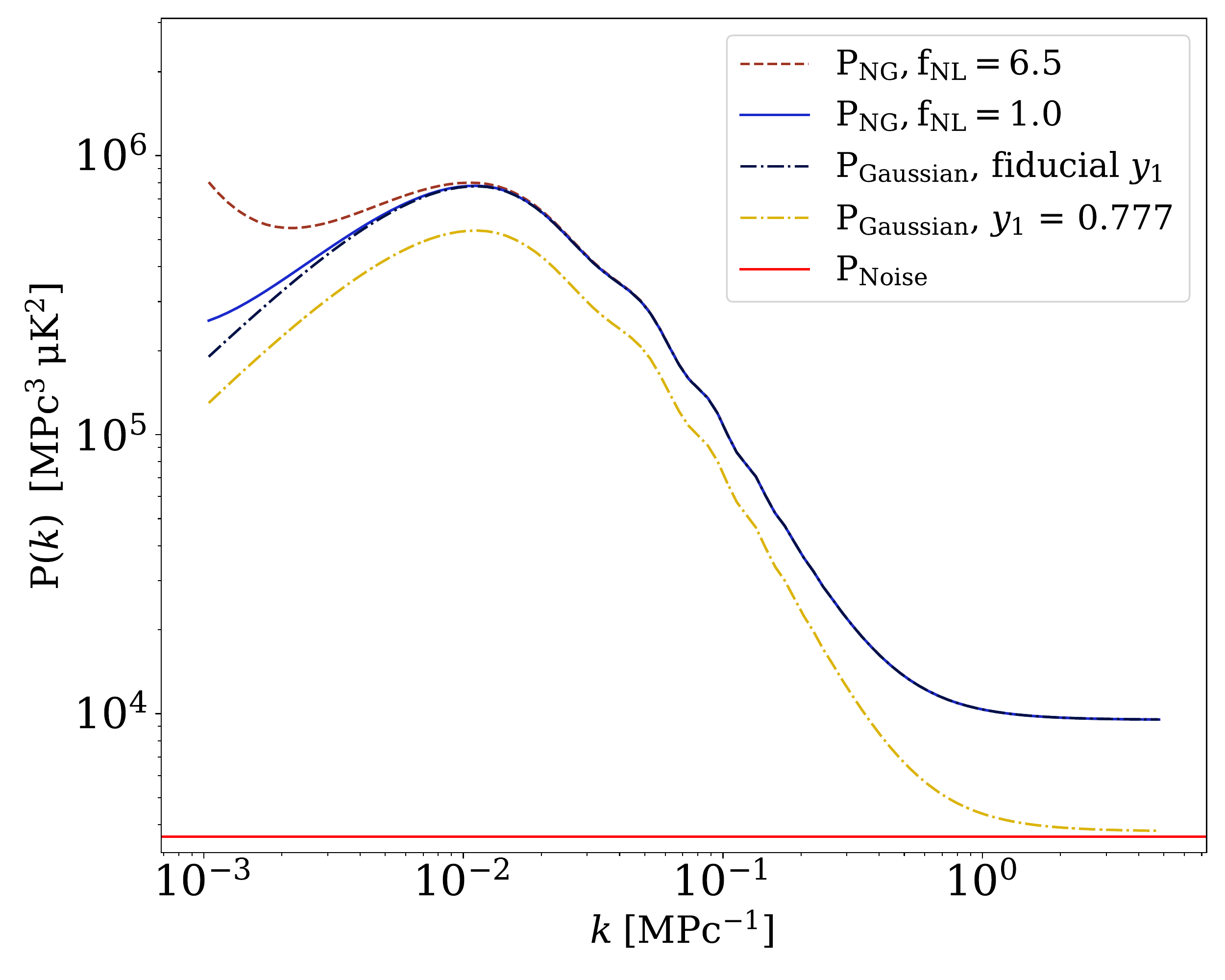}
    \caption{\pcb{The auto-power spectrum $P_1$ of our fiducial CO(1-0) model (thick blue curve) and its dependence on some of our model parameters.  The noise power spectrum of our hypothetical survey is plotted in red for comparison.}}
    \label{fig:Powerspectra}
\end{figure}

The model above is intended to predict $L(M)$ for the CO(1-0) line.  Rather than attempt to create a separate, highly uncertain model for CO(8-7), we will instead simply assume the same set of parameters for both lines, and examine how $\sigma_{\fnl}$ changes as we vary the CO(8-7) model.  For a `reasonable' range of parameter values, we will use the 2$\sigma$ range given by the \citet{Padmanabhan2018} parameter uncertainties.  For the remainder of this paper, $P_1$ will refer to the CO(1-0) power spectrum which we hold constant and $P_2$ will refer to the CO(8-7) power spectrum which we vary. \pcb{Figure \ref{fig:Powerspectra} shows the our model power spectrum, as well as the potential effects of the $\fnl$ and $y_1$ parameters.}

\begin{table}
    \centering
    \begin{tabularx}{0.75\columnwidth}{cc}
    \toprule
    Parameter $X$ & \makecell{Value at \\ $z=2.83$} \\
    \midrule
    \makecell{$A$ \\ $[\mathrm{K\; km\;s^{-1}\; pc^2}\;\Msun^{-1}]$} & $0.0328\pm 0.0222$ \\
    $M_1$  & $(1.76\pm 1.10)\times 10^{12}\;\Msun$ \\
    $y_1$ & $1.30\pm 0.53$ \\
    $y_2$ & $0.416\pm 0.366$ \\
    \bottomrule
    \end{tabularx}    
    \caption{Table of fiducial parameters used in the double power-law emission model. Uncertainties on these parameters and the fitting method are described in \citet{Padmanabhan2018}. We display the propagated uncertainties at $z=2.83$ for the convenience of the reader.}
    \label{tab:doublepowerlaw}
\end{table}

It should be noticed that our primary interest here is the multi-tracer effect, we are less interested in the overall signal-to-noise ratio attainable for any one experiment.  More detailed and specific forecasts can be found in \citet{Dizgah2019b}.  In order to isolate the multi-tracer impacts from overall signal-to-noise effects, we will always vary the amplitude $A$ of the CO(8-7) model at the same time as any of the other shape parameters in such a way that the mean intensity $\overline{T}$ remains constant.  \pcb{Under this assumption, we see in Fig. \ref{fig:Powerspectra} that the effect of, for example, decreasing $y_1$ is to slightly decrease the bias and more substantially decrease the shot noise.}

For both tracers, we utilise the Tinker mass function and bias models \citep{Tinker2008, Tinker2010}. The relevant matter power spectra $P_m(k)$ are computed using \texttt{CAMB} \citep{Lewis2000}.

\subsection{Survey Designs}
\label{sec:Survey_Design}

Here we describe our example survey designs. In order to focus on cosmic variance effects on the $\fnl$ constraints, we consider hypothetical near-future versions of the COMAP and CCAT-prime experiments. We consider the modified version of COMAP \citep{Li2016} to have a 10 m aperture with $N_{\mathrm{det}} = 1000$ detectors, observing over the current frequency range of 26-34 GHz (corresponding to $z=2.38-3.42$ for CO(1-0)). We further assume a total run time of $t_{\mathrm{obs}} = 2\times 10^4$ hours. A similar set of futuristic COMAP survey parameters were used in \citet{Dizgah2019a} to constrain $\sigma_{\fnl}$ to order unity. Other survey parameters are left unchanged from COMAP Phase 1, and are described in Table \ref{tab:surveyparams}.

In COMAP-type surveys, the noise power spectrum is defined as 
\begin{equation}
    P_N = \sigma_{\mathrm{vox}}^2 V_{\mathrm{vox}}
\end{equation}
Where $V_{\mathrm{vox}}$ is the volume of a single voxel, and
\begin{equation}
    \sigma_{\mathrm{vox}} = \dfrac{T_{\mathrm{sys}}}{\sqrt{N_{\mathrm{det}} \, \delta\nu \, t_{\mathrm{pix}}}}
\end{equation}
is defined as the noise in a single voxel. $T_{\mathrm{sys}}$ is the system temperature, $N_{\mathrm{det}}$ is the number of detectors, $\delta\nu$ is the frequency resolution and $t_{\mathrm{pix}}$ is observation time for each pixel, defined as $t_{\mathrm{pix}}= t_{\mathrm{obs}} \frac{\Omega_{\mathrm{pix}}}{\Omega_{\mathrm{surv}}} $. $\Omega_{\mathrm{pix}}$ and $\Omega_{\mathrm{surv}} $ are the angular size of each pixel and the size of the field, respectively. 

In order to cover the same redshift range as our CO(1-0) survey, we choose the frequency range of our future-CCAT to be 208-272 GHz. We somewhat arbitrarily choose with $N_{\mathrm{det}} = 1000$ detectors, and a total run time of $t_{\mathrm{obs}} = 2\times 10^4$ hours. As with the COMAP case, other survey parameters are left unchanged from the current version of CCAT prime, and are described in detail in Table \ref{tab:surveyparams}. 

In the case of CCAT-prime, the instrumental noise power spectrum is defined slightly differently by convention. Following \citet{Chung2018}, $P_N$ is defined as:
\begin{equation}
    P_N = \dfrac{\sigma_{\mathrm{pix}}^2} {t_{\mathrm{pix}}N_{\mathrm{det}}} V_{\mathrm{vox}}
\end{equation}
Here $V_{\mathrm{vox}}$, $t_{\mathrm{pix}}$ and $N_{\mathrm{det}}$ follow the same definitions as the COMAP case, while $\sigma_{\mathrm{pix}}$ is the on-sky sensitivity per sky pixel, specified in each survey. In CCAT prime, $\sigma_{\mathrm{pix}} = 0.86\; \mathrm{MJy/sr}\, \mathrm{s}^{1/2}$, taken from Table 1 of \citet{Chung2018}. 

We assume the fiducial version of both of these surveys cover the same 100 deg$^2$ field.  As we are mainly interested in very large-scale behavior here, we conservatively cut off our analyses at $k_{\rm{max}}=0.1$ Mpc$^{-1}$.  This lets us ignore impacts from non-linear power or resolution limits.  For the low-$k$ limit, we scale our observed power spectra by a window function set by the survey shape, as described in \citet{Bernal2019}.  

\begin{table}
    \centering
    \begin{tabularx}{\columnwidth}{@{\extracolsep{\fill}}lcc}
        \toprule
         & \makecell{Improved \\ COMAP} & \makecell{Improved \\ CCAT-p} \\
        \midrule
        System Temperature $T_{\mathrm{sys}}$ [K] & 40 & N/A \\[0.8ex]
        \makecell[l]{On-Sky Sensitivity Per Sky \\  Pixel $\sigma_{\mathrm{pix}}$ [$\mathrm{MJy}\;\mathrm{sr}^{-1}\;\mathrm{s}^{1/2}$]} & N/A &0.86 \\[2ex]
        Number of Detectors $N_{\mathrm{det}}$ & 1000 & 1000 \\
        Beam FWHM $\theta_{\mathrm{FWHM}}$ & 4' & 46.0" \\
        Frequency Range $\nu_{\mathrm{obs}}$ [GHz] & [26-34] & [208-272] \\
        Survey Area $\Omega_{\mathrm{surv}}$ $[\mathrm{deg}^2]$ & 100 & 100 \\
        Observing Time $t_{\mathrm{obs}}$ [hr] & $2\times 10^4$ & $2\times 10^4$ \\
        Channel Width $\delta\nu$ & 2 MHz & 2.5 GHz \\
        \bottomrule
    \end{tabularx}
    \caption{Table of survey parameters used in this work. Parameters not specified as changed in text are taken from current surveys, quoted in \citet{Breysse2019} (for COMAP) and \citet{Chung2018} (for CCAT-P)}
    \label{tab:surveyparams}
\end{table}

\section{Results}
\label{sec:Results}
We now explore how the $\sigma_{\fnl}$ depends on the various parameters of our example model.

\subsection{Effect of Emission Model}
\label{sec:ST_MT_Comparison}
\begin{figure*}
    \centering
    \includegraphics[width=2\columnwidth]{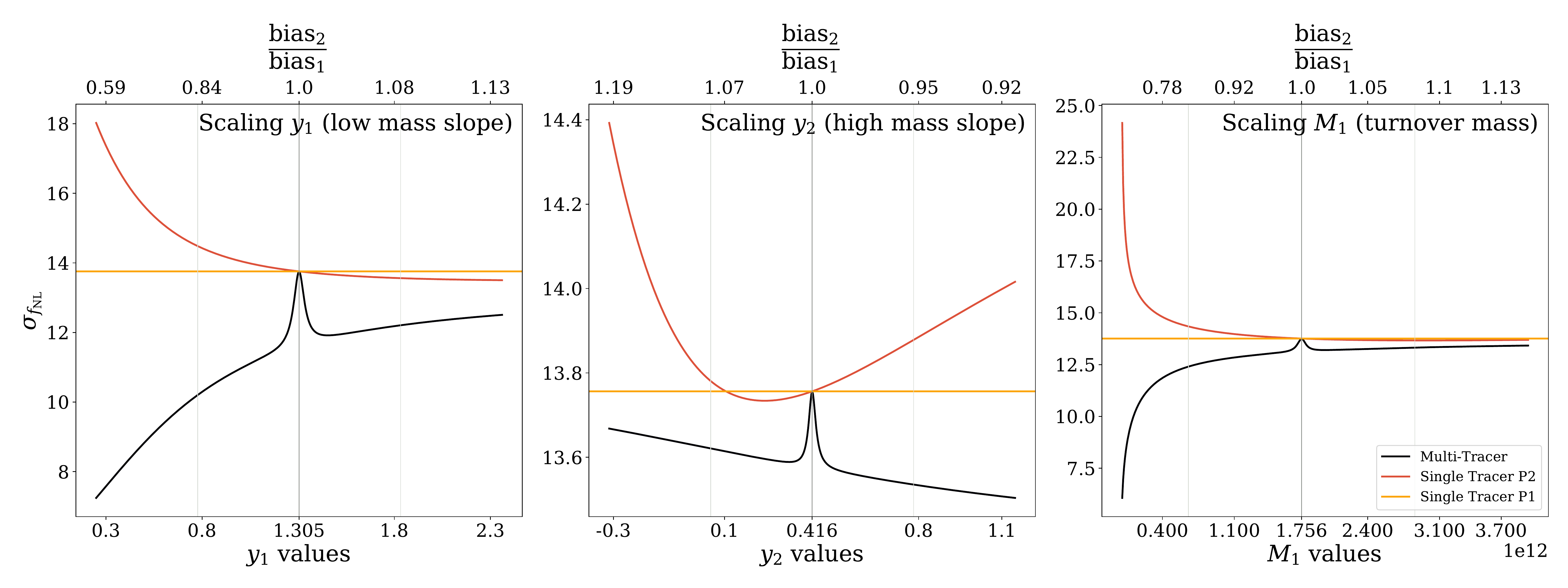}
    \caption{Comparison of Fisher forecasted $\fnl$ constraints utilising single and multiple tracers \pcb{in the sample variance limit}. The bottom axis shows the values of each of the varied parameters in the $P_2$ luminosity model, while the bottom axis shows the relative bias ratio between the $P_1$ and $P_2$ tracers. The darker grey line details the degeneracy point where the bias ratio equals one, while the lighter grey lines represent $\pm1\sigma$ deviations from the fiducial value, using uncertainties in Table \ref{tab:doublepowerlaw}. Observe that in each case, the multi-tracer constraint on $\fnl$ is bounded above by the single tracer constraints.}
    \label{fig:ST_MT_Plot}
\end{figure*}

Our arguments above show that multi-tracer effects should improve $\fnl$ measurements when the $L(M)$'s for our two different lines have different shapes.  \pcb{In order to gain intuition for the underlying physics, we work first in the sample variance-dominated limit by scaling the noise power spectrum of both surveys somewhat arbitrarily by 0.0005.  We will then go on to study the contribution from instrumental errors.  We choose to reduce the noise to a small number rather than zero as the covariance matrix becomes degenerate at zero noise.}  Figure \ref{fig:ST_MT_Plot} shows the results of a Fisher forecast on $\fnl$ \pcb{in the sample variance limit} varying the \pcb{shape parameters of the CO(8-7) $L(M)$ model}, comparing the multi-tracer analysis to the single-tracer results of each line.  Note again that we have chosen to hold the CO(1-0) model $P_1$ constant at the fiducial model parameters above, and that we have scaled the amplitude parameter $A$ of the CO(8-7) model to maintain constant $\overline{T}$.  All other CO(8-7) parameters are fixed at their fiducial values.

\pcb{The shapes of the curves in Fig. \ref{fig:ST_MT_Plot} are created by the interplay between the sample variance cancellation created by altering the bias and the simultaneous changes in the shot noise amplitude.} We observe from Figure \ref{fig:ST_MT_Plot} that for all CO(8-7) models, the multi-tracer Fisher forecast yields the same or lower $\sigmafnl$ values. We also see that the multi-tracer forecasts are heavily bias-dependent, and we obtain better constraints on $\fnl$ when the two tracers have very different bias. As expected, when $\frac{b_2}{b_1}=1$ the multi-tracer $\sigmafnl$ degenerates back to the single tracer case (see Appendix \ref{app:same_bias}). The single-tracer forecasts are also equal at this point, as our models for the two lines are identical and both noise spectra are negligible.  Our choice of a low-but-nonzero noise level sets the width of the peak near a bias ratio of 1.  Lower noise levels would lead to a narrower, sharper peak, becoming infinitely narrow at zero noise.

We see that out of the three model parameters we scaled in Figure \ref{fig:ST_MT_Plot}, the $y_1$ parameter, representing the $L(M)$ slope for low mass, had the largest effect on $\sigmafnl$.   We will thus focus on variation of $y_1$ alone for the remainder of this work, holding the other parameters at their fiducial values.

\subsection{Effect of Instrumental Noise}
\label{sec:Instrumental_Noise}

Now what we have seen the baseline dependence of $\sigma_{\fnl}$ on model parameters, we will explore the effects of instrument and survey design.  As shown in Section \ref{sec:Survey_Design}, there are a number of different parameters which affect survey sensitivity.  However, for our purposes, we are only interested in two properties: the overall amplitude of the noise power spectrum and the target survey volume.  These are the primary effects which determine the ratio of noise to sample variance uncertainty.  The $P_N$ amplitude effect is clear, decreasing $P_N$ increases the importance of sample variance.  For larger survey volumes, we gain access to lower-$k$ modes, which both have a more significant PNG contribution and more sample variance.  However, changing $\Omega_{\rm{surv}}$ also will have its own impacts on $P_N$ assuming fixed sensitivity and observing time.  

\begin{figure*}
    \centering
    \includegraphics[width=2\columnwidth]{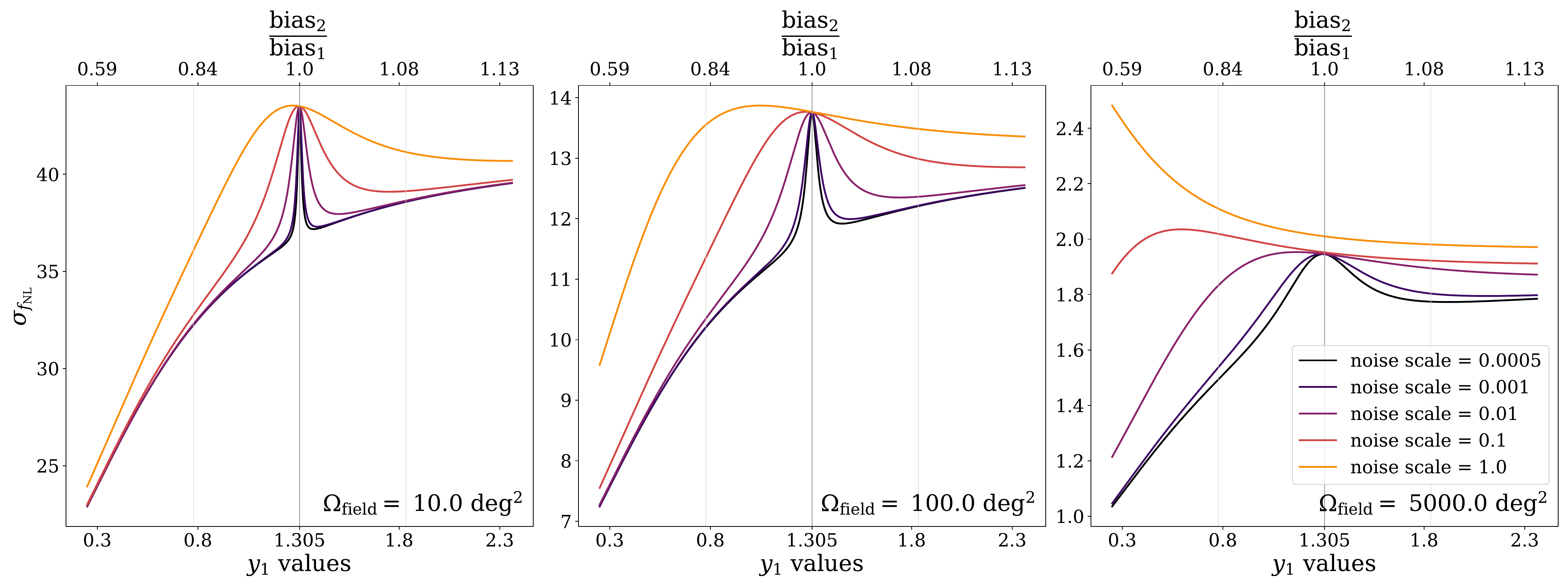}
    \caption{Bias ratio dependence of $\sigma_{\fnl}$ at different observation field sizes. A larger survey area yields a lower overall constraint on $\fnl$, but the dependence of $\sigmafnl$ on bias ratios is also much smaller. As above, the grey lines represent the fiducial parameter values and uncertainties from Table \ref{tab:doublepowerlaw}.}
    \label{fig:omega_field_plot}
\end{figure*}

Figure \ref{fig:omega_field_plot} shows the effects of both of these variations.  All three plots show the change in $\sigma_{\fnl}$ with low-mass slope.  We do not plot the other $L(M)$ parameters, but their qualitative behavior can be deduced by comparison to Fig. \ref{fig:ST_MT_Plot}.  The different coloured lines show the results of scaling $P_N$ from the above surveys by constant factors.  The central panel shows our fiducial 100 deg$^2$ survey, the others show the effect of changing survey area.  As expected, the sharp peak around the fiducial $y_1$ value indicative of multi-tracer cancellation is strongest when the survey is most dominated by sample variance.  Increasing the noise amplitude, whether by decreasing sensitivity or moving to larger survey area, reduces the impact of the multi-tracer effects.  As we move to larger survey areas, however, the overall sensitivity to $\fnl$ increases due to the addition of more low-$k$ modes.  Note that we have implicitly assumed that we are working with a flat sky, an assumption which breaks down at larger sky areas.  We leave a full curved-sky treatment for future work.

\subsection{Effects of Cross-Shot Power}
As described previously, a key distinction between LIM and galaxy survey measurements is the presence of a non-trivial cross-shot noise term, and its amplitude depends on how strongly correlated the two line luminosities are in a given galaxy.  We accounted for this correlation dependence by including a $s_\times$ parameter in Eq. \ref{eq:Pxshot}, and up to this point we have left $s_\times=1$.  We will now see what happens when we vary this quantity.  

\begin{figure}
    \centering
    \includegraphics[width=\columnwidth]{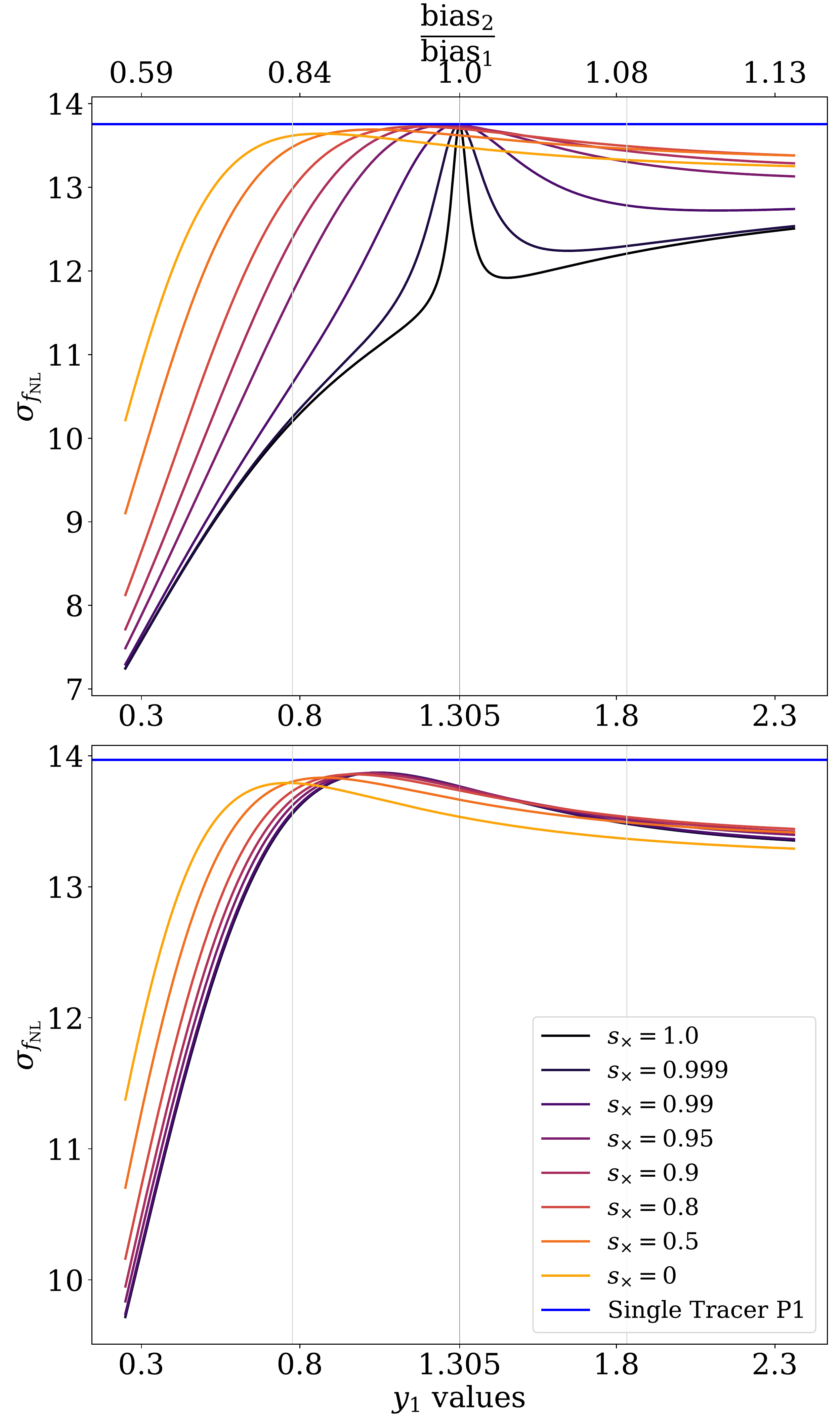}
    \caption{Fisher forecasted multi-tracer $\sigmafnl$ curves for different values of $s_\times$ in both the \pcb{sample variance} (top) and fiducial noise limits (bottom). We note the degeneracy peak at the same-bias point only occurs for high $s_\times$ values in the limit of high SNR. The multi-tracer $\sigmafnl$ curves are also bounded above by the single tracer $\sigmafnl$ value.}
    \label{fig:pxshotscale}
\end{figure}

We will again \pcb{begin} our forecasts in the high SNR limit, with the instrumental noise scaled to 0.0005 to remove any noise effects as described in Section \ref{sec:ST_MT_Comparison}. \pcb{The top panel of} Figure \ref{fig:pxshotscale} again shows the $y_1$-dependence of $\sigma_{\fnl}$ with different values of $s_\times$ for each $\sigmafnl$ curve.  The reduction or absence of cross-shot power changes the degeneracy structure of the covariance matrix from Eq. (\ref{eq:covariance_matrix}).  Even a small decrease in $s_\times$ substantially decreases the strength of the multi-tracer effect.  At worst, going from $s_\times=1$ to 0 worsens the $\fnl$ constraints by $\sim50\%$, going from $\sim9$ to $\sim14$.  \pcb{This may be counter-intuitive, as \emph{decreasing} a source of ``noise" \emph{improves} our measurement.  We can thus infer that the cross-correlation is cancelling out sample variance in the Poisson modes as well as clustering modes, an effect which does not exist in galaxy surveys.  The effect mostly disappears when we add back in instrument noise in the bottom panel of Figure \ref{fig:pxshotscale}, so it will likely not be important for near-future observations, at least under this model.}

\section{Discussion}
\label{sec:Discussion}

The shapes of the above multi-tracer forecasts are due to the interaction of a number of aspects of the survey design and $L(M)$ modeling.  As we have established, for very high SNR measurements, the multi-tracer case approaches the single when the two models are identical, or differ only in amplitude.  But the extra information in the multi-tracer modes comes in when we vary one of the models, leading to the sharp decrease in $\sigma_{\fnl}$ as we move away from the identical case.  The broader shapes are determined by the relative variation of the clustering and shot-noise components of the power spectrum, which depend on integrals over $L(M)$.  The details of these variations will be highly sensitive to exactly what $L(M)$ models are assumed.  A similar correlation between bias and shot noise exists in galaxy surveys, as more abundant populations will have lower shot noise and will tend to be less biased, but the coupling is more subtle, complex, and important in the intensity mapping case.

The addition of instrument noise further complicates this behavior.  Instrumental effects enter galaxy surveys by determining the faintest sources that can be detected, but since we are not attempting to detect individual galaxies, the instrument noise becomes an additional random field which is added to the map.  The impact of a multi-tracer analysis therefore depends on the relative amplitudes of the clustering signal, the intrinsic shot noise, and the noise power spectrum.

The presence of $\Pxshot$ in multi-tracer intensity mapping surveys has not yet been explored in much detail. \pcb{We have seen that $\Pxshot$ can have important impacts on $\sigma_{\fnl}$, though only at sensitivies higher than the near-future experiment we consider here.  However, between the unusual effects we see at high SNR and the relative paucity of references to this term in the literature, we argue that it will be important to take care to include the cross-shot noise term in future calculations.}  This is in line with other studies which have found important impacts of $\Pxshot$ on other aspects of LIM cross-correlations.  For example, \citet{Breysse2017} found that, because $\Pxshot$ depends differently on $L(M)$ than any component of the auto-spectrum, it adds substantially to attempts to study the very sub-galactic physics which complicate our analysis here.  A similar term also appears in correlations between galaxy surveys and intensity maps, allowing for more detailed understanding of the connection between detected galaxies and unresolved emission \citep{Wolz2017,Wolz2019,Breysse2019}.

It is important to note that our treatment of stochasticity here is highly approximate, intended to be more illustrative than quantitative.  As seen in \citet{Seljak2009}, stochasticity can have a substantial impact on every aspect of the general multi-tracer problem.  The addition of interstellar medium physics makes this stochasticity difficult to model.  There is some evidence that there is significant scatter around a mean $L(M)$ relation \citep{Li2016}, but we have very little idea now how large that scatter will be, and even less about how much we expect the line \emph{ratios} in individual galaxies to vary.  And again, we have found that even a relatively small variation \pcb{can} matter.  We leave a full treatment of stochasticity effects for future work.

There are several other approximations we have made here to focus on the specific effects we wanted to illustrate.  Our focus was limited to $\fnl$ measurements, a full analysis would need to simultaneously fit for the full range of astrophysical and cosmological parameters, which may introduce degeneracies.  We neglected effects of anisotropy in the power spectrum, which would further complicate the analysis \citep{Dizgah2019b,Bernal2019,Chung2019}.  Finally, we did not include any foreground contamination in our forecasts.  The impact of foregrounds on intensity mapping varies wildly depending on which line is being considered.  CO(1-0) surveys are only mildly contaminated \citep{Chung2017}, but the CO(8-7) map would have to contend with lower-order CO transitions coming from lower redshifts.  It is possible that this contamination could be removed, either by masking out contaminated regions \citep{Gong2012,Sun2018}, or by exploiting power spectrum anisotropies \citep{Lidz2016,Cheng2016}.  The problem is even worse for 21 cm surveys, as diffuse emission from the Milky Way is often 3-5 orders of magnitude brighter than the cosmological signal \citep{Oh2003,Wang2006,Liu2011}.  Since we are proposing rather futuristic measurements here, it is likely that these foreground problems will be dealt with by the time experiments reach our described sensitivity, but this may still be an important consideration for future studies.

We have presented here an example of a single, hypothetical cross-correlation, but CO(1-0)$\times$CO(8-7) is far from the only possible multi-tracer measurement, and we make no claim that it is necessarily the best or easiest.  There are already many LIM cross-correlations under consideration, and more will likely appear as more data become available.  Balloon-borne surveys like EXCLAIM can access the bright CII line at the same $z\sim3$ redshift range we consider here \citep{Padmanabhan2019}.  The SPHEREx survey will map a variety of lines over the whole sky \citep{Gong2017}.  If any of these lines have different biases, then SPHEREx or its successors could make this type of measurement through internal cross-correlations.  The multi-tracer correlation between SPHEREx H$\alpha$ and the 21 cm line has already seen some study \citep{Fonseca2018}, though without some of the effects we considered here.

\section{Conclusions}
\label{sec:Conclusions}

Every aspect of intensity mapping surveys is intimately coupled to the physics of line emission.  We have shown here that measurements of the PNG feature in the power spectrum, an effect which is most dominant on Gpc or greater scales, can depend sensitively on pc-scale ISM properties.  Intensity maps, with their ability to cover large scales relatively quickly and cheaply, are well-suited to this type of PNG observation, but our work demonstrates that one must carefully account for the details of this astrophysical coupling when making forecasts.  Sub-galactic physics determines the overall signal-to-noise of a LIM measurement, but also determines the average bias of a map, and sets the shot noise level in both the auto- and cross-spectra, all effects which we have shown to be important for $\fnl$.  In the model we studied here, we found that including these effects can change forecasted $\fnl$ constraints by order unity.

Our example CO correlation illustrates some important aspects of these effects, but it is far from the full picture.  Our work demonstrates that, as early LIM surveys begin to produce maps, there is value in identifying lines with very different biases.  We find that the greatest benefits can be found by identifying lines with diverging behavior at lower halo masses.  First-generation LIM surveys, due to their sensitivity at low luminosities, are uniquely suited to identifying candidate lines which show this behavior.  If lines with different biases can be identified, we have shown that a cross-correlation between them would provide powerful benefits as a probe of primordial physics.

Intensity mapping remains a young field.  Many of the ways in which it differs from other cosmological observables are only beginning to be explored.  Detailed studies like this one of the unique coupling between galactic and cosmological scales will be critical as we seek to reach the full potential of these exciting measurements.

\begin{acknowledgments}
We would like to acknowledge the Natural Sciences and Engineering Research Council of Canada for making this research possible.

\end{acknowledgments}

\appendix

\section{Cross Power Spectra Derivation}
\label{app:Px}

Consider two intensity map observables, $T_1$ and $T_2$. We divide the map into $N_c$ infinitesimal cells, and we consider only the intensity contributed by sources with a CO luminosity between $L$ and $L+dL$. The infinitesimal cells are defined small enough such that the number of source emitters within each cell is either zero or 1.
The intensity in cell $i$ from these sources is 
\begin{equation}
    T(\mathbf{x}_i, L)\,dL = C_{LT}L\,N_i(L)\,dL
\end{equation}
Where $N_i(L)=0\text{ or }1$ is the number of emitters within the cell, while $C_{LT}$ is the conversion factor as defined in Equation \ref{eq:CLT}.

The intensity mapping cross-power spectrum $P_{\times}(k)$ is defined in a similar manner to the auto power spectrum:
\begin{equation}
    P_{\times}(k) \equiv V \Big\langle \Tilde{T}_1(\mathbf{k})\;\Tilde{T}_2^*(\mathbf{k})\Big\rangle
\end{equation}
where the tilde indicates a Fourier Transform
\begin{align}
\begin{split}
    \Tilde{T}(\mathbf{k}, L)\,dL &= \frac{1}{V} \int T(\mathbf{x}, L) e^{i\mathbf{k}\cdot\mathbf{x}} d^3x \\
    &= C_{LT}\frac{L}{V} \sum_{i=1}^{N_c} N_i(L)\; e^{i\mathbf{k}\cdot\mathbf{x_i}}\, dL
\end{split}
\end{align}

Next, we split the contributions to $P_\times(k)$ into first contributions from different cells ($i\neq j$) and then contributions from the same cells. We also consider sources with different luminosities. Starting with the two-source part, we have
\begin{align}
\begin{split}
    \Big\langle \Tilde{T}_1(\mathbf{k}, L_1)\;\Tilde{T}_2^*(\mathbf{k}, L_2)&\Big\rangle_{i\neq j} dL_1 dL_2 = C_{LT}^2\frac{L_1L_2}{V^2}\\ 
    & \times \sum_{i\neq j} \langle N_{i}(L_1)N_{j}(L_2)\rangle e^{i\mathbf{k}\cdot(\mathbf{x_{i}}-\mathbf{x_{j}})}dL_1\, dL_2
\end{split}
\end{align}
If we assume mass-luminosity relations for our two target lines, then $L N(L)dL=L(M)N(M)dM$, and we have
\begin{align}
\begin{split}
\Big\langle \Tilde{T}_1(&\mathbf{k}, M_1)\;\Tilde{T}_2^*(\mathbf{k}, M_2)\Big\rangle_{i\neq j} dM_1 dM_2=\\
&C_{LT}^2\frac{L_1(M)L_2(M)}{V^2} \times \sum_{i\neq j}\langle N_{i}(M_1)N_{j}(M_2)\rangle e^{i\mathbf{k}\cdot(\mathbf{x_{i}}-\mathbf{x_{j}})}dM_1\, dM_2
\end{split}
\end{align}

The expectation value $\left<N_i(M)\right>=dn/dM \delta V$.  With this and an assumption of a linear bias $b(M)$ we obtain
\begin{align}
\begin{split}
\Big\langle &\Tilde{T}_1(\mathbf{k}, M_1)\;\Tilde{T}_2^*(\mathbf{k}, M_2)\Big\rangle_{i\neq j} dM_1 dM_2\\ &=\frac{C_{LT}^2}{V^2}L_1(M_1)\frac{dn}{dM_1}L_2(M_2)\frac{dn}{dM_2}\\ &\times\sum_{i\neq j}\delta V^2\left[1+b(M_1)b(M_2)\xi_m(\mathbf{x}_i-\mathbf{x}_i)\right]e^{i\mathbf{k}\cdot(\mathbf{x_{i}}-\mathbf{x_{j}})}dM_1\, dM_2
\end{split}
\end{align}
where $\xi_m$ is the matter two-point correlation function.  Carrying out the Fourier transform yields
\begin{multline}
\Big\langle\Tilde{T}_1(\mathbf{k}, M_1)\;\Tilde{T}_2^*(\mathbf{k}, M_2)\Big\rangle_{i\neq j} dM_1 dM_2 = \\ \frac{C_{LT}^2}{V}L_1(M_1)b(M_1)\frac{dn}{dM_1}L_2(M_2)b(M_2)\frac{dn}{dM_2}P_m(k)
\end{multline}
We can then integrate over all possible halo masses to get
\be
\Big\langle\Tilde{T}_1(\mathbf{k}, M_1)\;\Tilde{T}_2^*(\mathbf{k}, M_2)\Big\rangle_{i\neq j}=\frac{1}{V}\overline{T}_1\overline{T}_2\overline{b}_1(k)\overline{b}_2(k)P_m(k)
\label{eq:Pclustx_deriv}
\ee

We turn now to the term where $i=j$, which encodes the correlation of each source with itself.  With our mass-luminosity models assumed, we have
\begin{align}
\begin{split}
    \Big\langle \Tilde{T}_1(&\mathbf{k}, M_1)\;\Tilde{T}_2^*(\mathbf{k}, M_2)\Big\rangle_{i= j} dM_1 dM_2 = \\ 
    &C_{LT}^2\frac{L_1(M_1)L_2(M_2)}{V^2} \times \sum_{i}^{N_c} \langle N_{i}(M_1)N_{i}(M_2)\rangle e^{i\mathbf{k}\cdot(\mathbf{x_{i}}-\mathbf{x_{j}})}dL_1\, dL_2
\end{split}
\end{align}
Continuing to take advantage of the properties of $\left<N_i\right>$, we can write
\begin{align}
\begin{split}
\left<N_i(M_1)N_i(M_2)\right>&=\left<N_i(M_1)\right>\delta_D(M_1-M_2) \\
&=\frac{dn}{dM_1}\delta V\delta_D(M_1-M_2)
\end{split}
\end{align}
where the Dirac delta function enforces that $M_1$ and $M_2$ must be the same if both refer to the same halo.  We can then write
\begin{multline}
\Big\langle\Tilde{T}_1(\mathbf{k}, M_1)\;\Tilde{T}_2^*(\mathbf{k}, M_2)\Big\rangle_{i= j}dM_1dM_2=\frac{C_{LT}^2}{V^2}L_1(M_1)L_2(M_2)\frac{dn}{dM_1}\\ \times\delta_D(M_1-M_2)\sum_{i=1}^{N_c}\delta Ve^{i\mathbf{k}\cdot\mathbf{x}}dM_1dM_2
\end{multline}
Integrating over $M_1$ and $M_2$ then leaves
\be
\Big\langle\Tilde{T}_1(\mathbf{k}, M_1)\;\Tilde{T}_2^*(\mathbf{k}, M_2)\Big\rangle_{i= j}=\frac{1}{V}P^{\rm{shot}}_\times.
\label{eq:Pshotx_deriv}
\ee
Combining Eqs. (\ref{eq:Pclustx_deriv}) and (\ref{eq:Pshotx_deriv}) yields our full form for the cross-power spectrum:
\be
P_\times(k)=\overline{T}_1\overline{T}_2\overline{b}_1(k)\overline{b}_2(k)P_m(k)+P^{\rm{shot}}_\times.
\ee

\section{Same-bias Case}
\label{app:same_bias}

Here we will demonstrate that the multi-tracer analysis loses much of its advantages in the case where the two lines have the same bias.  Note that our definition of the intensity mapping bias is independent of the overall normalization of $L(M)$, so we can have two lines with different average intensities that none the less have the same bias.  Let us consider this case by defining
\be
L_2(M)\equiv AL_1(M).
\ee

First, let us consider the case where we have only one line, and we do not have access to the multi-tracer method.  In this case, our covariance matrix $C$ only has one entry, $P_1(k)+P_N$, and the single-tracer Fisher matrix simply becomes
\be
F_{\fnl\fnl}^{\rm{ST}}=\sum_k\frac{N_m(k)}{2}\left(\frac{\partial P_1(k)}{\partial\fnl}\right)^2\frac{1}{\left(P_1(k)+P_N\right)^2}.
\ee
We can easily see how this would simplify in the two extreme limits, where the survey is noise-dominated,
\be
F_{\fnl\fnl}^{\rm{ST}}(P_N\gg P_1)=\sum_k\frac{N_m(k)}{2}\left(\frac{\partial P_1(k)}{\partial\fnl}\right)^2\frac{1}{P_N^2},
\ee
and in the sample variance limit,
\be
F_{\fnl\fnl}^{\rm{ST}}(P_1\gg P_N)=\sum_k\frac{N_m(k)}{2}\left(\frac{\partial P_1(k)}{\partial\fnl}\right)^2\frac{1}{P_1^2(k)}.
\ee

Now we turn to the multi-tracer case.  Under our assumption that $L_1(M)$ and $L_2(M)$ have the same slope, and taking the $s_\times$ parameter in $\Pxshot$ to be 1, we have:
\begin{equation}
\label{eq:Px_degeneracy}
\begin{split}
    P_\times(k) &= A\Bar{T}_1^2\Bar{b}_1^2(k)P_m(k) + AP_1^{\mathrm{shot}} \\
    &= AP_1(k)
\end{split}
\end{equation}
with the simplified assumption that $C_{LT,1}=C_{LT,2}$. Further, we see under our assumptions, $P_2(k) = A^2P_1(k)$. 
Our covariance matrix then becomes
\begin{align}
\label{eq:Degen_Covariance}
\begin{split}
    &C(k) = 
    \begin{pmatrix} 
    P_1(k) & P_\times(k) \\
    P_\times(k) & P_2(k) 
    \end{pmatrix} \\
    &= \begin{pmatrix} 
        P_1(k)+P_N & 
        AP_1(k) \\
        AP_1(k) & 
        A^2P_1(k)+P_N
        \end{pmatrix},
\end{split}
\end{align}
where we have assumed for simplicity that both surveys have the same noise.  Running this through the Fisher formula yields
\be
F_{\fnl\fnl}=\sum_k\frac{N_m(k)}{2}\frac{(1+A^2)^2}{\left[(1+A^2)P_1(k)+P_N\right]^2}\left(\frac{\partial P_1(k)}{\partial\fnl}\right)^2.
\ee

Let us consider again the two noise limits.  In the noise-dominated case, this reduces to
\be
F_{\fnl\fnl}(P_N\gg P_1)=\sum_k\frac{N_m(k)}{2}\frac{(1+A^2)^2}{P_N^2}\left(\frac{\partial P_1(k)}{\partial\fnl}\right)^2.
\ee
If the two lines are identical, i.e. if $A=1$, we get
\begin{align}
\begin{split}
F_{\fnl\fnl}(P_N\gg P_1)&=\sum_k\frac{N_m(k)}{2}\frac{4}{P_N^2}\left(\frac{\partial P_1(k)}{\partial\fnl}\right)^2 \\
&=4F_{\fnl\fnl}^{\rm{ST}}(P_N\gg P_1).
\end{split}
\end{align}
In other words,
\be
\sigma_{\fnl}(P_N\gg P_1)=\frac{1}{2}\sigma_{\fnl}^{\rm{ST}}(P_N\gg P_1).
\ee
This makes intuitive sense.  We get a factor of two improvement from having two different noise realizations, but no additional improvement beyond that.  

In the sample variance limit, we have
\begin{align}
\begin{split}
\label{eq:VarianceLimitDegen}
F_{\fnl\fnl}(P_1\gg P_N)&=\sum_k\frac{N_m(k)}{2}\left(\frac{\partial P_1(k)}{\partial\fnl}\right)^2\frac{1}{P_1^2(k)}\\ &=F_{\fnl\fnl}^{\rm{ST}}(P_1\gg P_N).
\end{split}
\end{align}
In other words, if we are sample-variance limited and the two lines differ only in amplitude then the second map adds no additional cosmological information.  Intuitively, it cannot add anything because the second map will only be the first multiplied by a constant.  This is why most of our constraints above converge to the single tracer case when $b_1=b_2$.

Furthermore, we note that the degeneracy described in the sample variance limit only exists in the case with $b_1=b_2$. We prove this by showing that for $b_1\neq b_2$, it is not possible for Equation \ref{eq:Px_degeneracy} to be satisfied, and thus Equations \ref{eq:Degen_Covariance} to \ref{eq:VarianceLimitDegen} would not apply. 

Observe that by the definition of halo bias in intensity mapping given by Equation \ref{eq:LIMbias}, $b_1=b_2$ if and only if $L_1(M)=L_2(M)$.
We also note that for Equation \ref{eq:Px_degeneracy} to be satisfied, the cross shot noise spectrum $\Pxshot$ must be a scalar multiple of $\Pshot_1$
\begin{equation}
\label{eq:SwartzEquality}
    P_\times^\mathrm{shot}=AP_1^{\mathrm{shot}} =\left(P_1^{\mathrm{shot}}P_2^{\mathrm{shot}}\right)^{\frac{1}{2}}
\end{equation} 
However, from Equations \ref{eq:Pshot} and \ref{eq:Pxshot} defining the auto and cross shot noise, we see
\begin{align}
\begin{split}
    \left(\Pxshot\right)^2 &= \left(s_\times C_{LT,1}C_{LT,2}\int L_1(M)\,L_2(M)\dv{n}{M} \dd M\right)^2 \\
    &\leq \left(C_{LT,1}^2 \int L_1(M)^2 \dv{n}{M} dM\right)\\
    & \phantom{aaaaaaa} \times \left(C_{LT,2}^2 \int L_2(M)^2 \dv{n}{M} dM\right) \\
    &= \Pshot_1 \Pshot_2
\end{split}
\end{align}
Through application of the Schwarz inequality. Furthermore, the Schwarz Inequality states that the two sides are only equal in the case where the luminosity models are linearly dependent, when $L_2(M)=AL_1(M)$.

Thus we see that in all other luminosity model pairs between the two tracers, $b_1\neq b_2$ and Equation \ref{eq:SwartzEquality} becomes Equation \ref{eq:SwartzInEquality}
\begin{equation}
\begin{split}
    \label{eq:SwartzInEquality}
    \left(\Pxshot \right)^2 < \Pshot_1 \Pshot_2 \\
    P_\times^\mathrm{shot} < \left(P_1^{\mathrm{shot}}P_2^{\mathrm{shot}}\right)^{\frac{1}{2}}
\end{split}
\end{equation}
And would thus not yield the type of degeneracy derived here and shown Figures \ref{fig:ST_MT_Plot} through \ref{fig:pxshotscale}. As a result, we can conclude that measuring multiple tracers with different astrophysical $L(M)$ would always yield a lower $\sigmafnl$ than measurements with a single tracer, and equality between multi-tracer and single-tracer measurements only occurs for lines with similar bias and luminosity models $L(M)$.


\bibliography{main}

\end{document}